\documentclass{ws-p8-50x6-00}
\newcommand{\ud}{\mathrm{d}}

\begin{document}

\title{LOCALIZED COHERENT STRUCTURES AND PATTERNS FORMATION IN COLLECTIVE
MODELS OF BEAM MOTION}
\author{A. Fedorova, M. Zeitlin}
\address {IPME, RAS, St.~Petersburg, 
V.O. Bolshoj pr., 61, 199178, Russia\\
e-mail: zeitlin@math.ipme.ru\\
http://www.ipme.ru/zeitlin.html;
http://www.ipme.nw.ru/zeitlin.html}
\maketitle

\abstracts{
We present applications of variational -- wavelet approach to 
three different models of nonlinear beam motions with underlying collective behaviour:
Vlasov-Maxwell-Poisson systems, envelope dynamics, beam-beam model.
We have the representation for dynamical variables as
a multiresolution (multiscales) expansion via high-localized nonlinear eigenmodes in
the base of compactly supported wavelet bases.
Numerical modelling demonstrates formation of coherent structures and stable patterns.
}

\section{Introduction}
In this paper we consider the applications of a new nu\-me\-ri\-cal\--analytical 
technique which is based on the methods of local nonlinear harmonic
analysis or wavelet analysis to three nonlinear beam/accelerator physics
problems which can be characterized by  collective type behaviour: 
some forms of Vlasov-Maxwell-Poisson equations[1], RMS envelope dynamics[2], the model of 
beam-beam interactions[3].
Such approach may be useful in all models in which  it is 
possible and reasonable to reduce all complicated problems related with 
statistical distributions to the problems described 
by systems of nonlinear ordinary/partial differential 
equations with or without some (functional)constraints.
Wavelet analysis is a relatively novel set of mathematical
methods, which gives us the possibility to work with well-localized bases in
functional spaces and gives the maximum sparse forms for the general 
type of operators (differential, integral, pseudodifferential) in such bases. 
Our approach is based on the 
variational-wavelet approach from [4]-[14],
which allows us to consider polynomial and rational type of 
nonlinearities.
The solution has the following multiscale/multiresolution decomposition via 
nonlinear high-localized eigenmodes 
\begin{eqnarray}\label{eq:z}
u(t,x)&=&\sum_{(i,j)\in Z^2}a_{ij}U^i(x)V^j(t),\\
V^k(t)&=&V_N^{k,slow}(t)+\sum_{i\geq N}V^k_i(\omega^1_it), \quad \omega^1_i\sim 2^i \\
U^k(x)&=&U_M^{k,slow}(x)+\sum_{j\geq M}U^k_j(\omega^2_jx), \quad \omega^2_j\sim 2^j, 
\end{eqnarray}
which corresponds to the full multiresolution expansion in all underlying time/space 
scales ($x$ are the generalized space coordinates or phase space coordinates, $t$ is time coordinate).
Formula (\ref{eq:z}) gives us expansion into the slow part $u_{N,M}^{slow}$
and fast oscillating parts for arbitrary N, M.  So, we may move
from coarse scales of resolution to the 
finest one for obtaining more detailed information about our dynamical process.
The first terms in the RHS of formulae (1)-(3) correspond on the global level
of function space decomposition to  resolution space and the second ones
to detail space. In this way we give contribution to our full solution
from each scale of resolution or each time/space scale or from each nonlinear eigenmode
(Fig.1). 
The same is correct for the contribution to power spectral density
(energy spectrum): we can take into account contributions from each
level/scale of resolution.
In all these models numerical modelling demonstrates the appearence of coherent high-localized structures
and stable patterns formation.
Starting  in part 2 from Vlasov-Maxwell-Poisson equations, root-mean-square (RMS) envelope dynamics
and beam-beam interaction model
we consider in part 3 the approach based on
variational-wavelet formulation. 
We give explicit representation for all dynamical variables in the base of
compactly supported wavelets or nonlinear eigenmodes.  Our solutions
are parametrized
by solutions of a number of reduced algebraical problems one from which
is nonlinear with the same degree of nonlinearity and the rest  are
the linear problems which correspond to particular
method of calculation of scalar products of functions from wavelet bases
and their derivatives. 
In part 4 we consider numerical modelling based on our analytical approach.
\begin{figure}[htb]
\epsfxsize=15pc
\figurebox{20pc}{20pc}{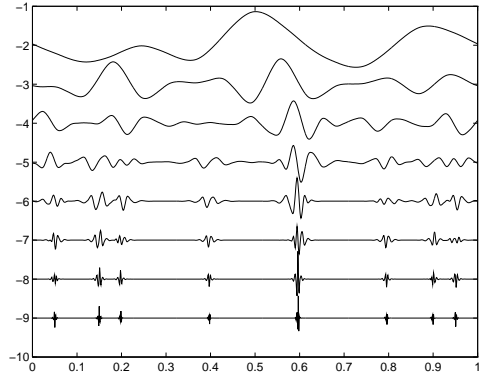}
\caption{Multiscale/eigenmode decomposition.}
\end{figure}

\section{Collective models}
\subsection{Vlasov-Maxwell-Poisson Equations}

Analysis based on the non-linear Vlasov-Maxwell-Poisson equations leads to more
clear understanding of the collecti\-ve effects and nonlinear beam dynamics
of high intensity beam propagation in periodic-focusing and uniform-focusing
transport systems.
We consider the following form of equations ([1],[2] for setup and designation): 
\begin{eqnarray}
&&\Big\{\frac{\partial}{\partial s}+p_x\frac{\partial}{\partial x}+
             p_y\frac{\partial}{\partial y}-
\Big[k_x(s)x+\frac{\partial\psi}{\partial x}\Big]\frac{\partial}{\partial p_x}-\nonumber\\
&& \Big[k_y(s)y+\frac{\partial\psi}{\partial y}\Big]\frac{\partial}{\partial p_y}
  \Big\} f_b(x,y,p_x,p_y,s)=0, \\
&&\Big(\frac{\partial^2}{\partial x^2}+\frac{\partial^2}{\partial y^2}\Big)\psi=
-\frac{2\pi K_b}{N_b}\int \ud p_x \ud p_y f_b,\\
&&\int\ud x\ud y\ud p_x\ud p_y f_b=N_b
\end{eqnarray}
The corresponding Hamiltonian for transverse sing\-le\--par\-ticle motion is given by 
\begin{eqnarray}
&& H(x,y,p_x,p_y,s)=\frac{1}{2}(p_x^2+p_y^2) 
                   +\frac{1}{2}[k_x(s)x^2 \\
 &&+k_y(s)y^2]+
    H_1(x,y,p_x,p_y,s)+\psi(x,y,s), \nonumber
\end{eqnarray}
where $H_1$ is nonlinear (polynomial/rational) part of the full Hamiltonian.
In case of Vlasov-Maxwell-Poisson system we may transform (4) into invariant form
 \begin{eqnarray}
\frac{\partial f_b}{\partial s}+[f,H]=0.
\end{eqnarray}

\subsection{RMS Equations}

We  consider an approach based on 
the second moments of the distribution functions for  the calculation
of evolution of RMS envelope of a beam.
The RMS envelope equations are the most useful for analysis of the 
beam self--forces (space--charge) effects and also 
allow to consider  both transverse and longitudinal dynamics of
space-charge-dominated relativistic high--brightness
axisymmetric/asymmetric beams, which under short laser pulse--driven
radio-frequency photoinjectors have fast transition from nonrelativistic
to relativistic regime [2]. Analysis of halo growth in beams, appeared
as result of bunch oscillations in the particle-core model, also are based
on three-dimensional envelope equations [2]. 
We can consider the different forms of RMS envelope equations,
which are 
not more than nonlinear differential equations with rational
nonlinearities and variable coefficients from the formal  point of view.
Let $f(x_1,x_2)$ be the distribution function which gives full information
 about 
noninteracting ensemble of beam particles regarding to trace space or 
transverse phase coordinates $(x_1,x_2)$. 
Then we may extract the first nontrivial effects of collective dynamics from 
the second moments
\begin{eqnarray}
\sigma_{x_i x_j}^2&=&<x_i x_j>=\int\int x_i x_j f(x_i,x_j)\ud x_i\ud x_j 
\end{eqnarray}
RMS emittance ellipse is given by
$
\varepsilon^2_{x,rms}=<x_i^2><x_j^2>-<x_i x_j>^2 (i\ne j)
$.
 Expressions for twiss  parameters are also based on the second moments.
We will consider the following particular
cases of RMS envelope equations, which describe evolution
of the moments (9) ([2] for full designation):
for asymmetric beams we have the system of two envelope equations
of the second order for $\sigma_{x_1}$ and $\sigma_{x_2}$:
\begin{eqnarray}
&&\sigma^{''}_{x_1}+\sigma^{'}_{x_1}\frac{\gamma '}{\gamma}+
\Omega^2_{x_1}\left(\frac{\gamma '}{\gamma}\right)^2\sigma_{x_1}=
{I}/({I_0(\sigma_{x_1}+\sigma_{x_2})\gamma^3})
+
\varepsilon^2_{nx_1}/{\sigma_{x_1}^3\gamma^2},\\
&&\sigma^{''}_{x_2}+\sigma^{'}_{x_2}\frac{\gamma '}{\gamma}+
\Omega^2_{x_2}\left(\frac{\gamma '}{\gamma}\right)^2\sigma_{x_2}= 
{I}/({I_0(\sigma_{x_1}+\sigma_{x_2})\gamma^3})
+
\varepsilon^2_{nx_2}/{\sigma_{x_2}^3\gamma^2}\nonumber
\end{eqnarray}
The envelope equation for an axisymmetric beam is
a particular case of preceding equations.
Also we have related Lawson's equation for evolution of the rms
envelope in the paraxial limit, which governs evolution of cylindrical
symmetric envelope under external linear focusing channel
of strenght $K_r$:
\begin{equation}
\sigma^{''}+\sigma^{'}\left(\frac{\gamma '}{\beta^2\gamma}\right)+
K_r\sigma=\frac{k_s}{\sigma\beta^3\gamma^3}+
\frac{\varepsilon^2_n}{\sigma^3\beta^2\gamma^2},\nonumber
\end{equation}
where
$
K_r\equiv -F_r/r\beta^2\gamma mc^2, \ \ \
 \beta\equiv \nu_b/c=\sqrt{1-\gamma^{-2}}.
$
According [2] we have the following form for envelope equations in the model of halo formation
by bunch oscillations:
\begin{eqnarray}
\ddot{X}+k_x^2(s)X-\frac{3K}{8}\frac{\xi_x}{YZ}-\frac{\varepsilon^2_x}{X^3}&=&0,\nonumber\\
\ddot{Y}+k_y^2(s)Y-\frac{3K}{8}\frac{\xi_y}{XZ}-\frac{\varepsilon^2_y}{Y^3}&=&0,\\
\ddot{Z}+k_z^2(s)Z-\gamma^2\frac{3K}{8}\frac{\xi_z}{XY}-\frac{\varepsilon^2_z}{Z^3}&=&0,\nonumber
\end{eqnarray}
where X(s), Y(s), Z(s) are bunch envelopes, $\xi_x, \xi_y$, $\xi_z= F(X,Y,Z)$.

After transformations to Cauchy form we can see that
all these equations from the formal point of view are not more than
ordinary differential equations with rational nonlinearities
and variable coefficients
Also, we may consider regimes in which $\gamma$, $\gamma'$
are not fixed functions/constants but satisfy some additional differential 
constraints/equations,
but this case does not change our general approach of the next part. 

\subsection{Beam-beam modelling}

In A.~Chao e.a. model [3] for simulation of beam-beam interaction the initial 
collective description by some sort of equation for distribution function $f(s,x,p)$
\begin{eqnarray}
\frac{\partial f}{\partial s}+p\frac{\partial f}{\partial x}-
  \large(k(s)x-F(x,s)\large)\frac{\partial f}{\partial p}=0
\end{eqnarray}
is reduced to Fockker-Planck (FP) equation on the first stage and to
very nontrivial dynamical system with complex behaviour,
\begin{eqnarray}
&&\frac{\ud^2\sigma_k}{\ud s^2}+ \Gamma_k\frac{\ud\sigma_k}{\ud s}+
 F_k\sigma_k=\frac{1}{\beta^2_ka^2_k\sigma^3_k}\nonumber\\
&&\frac{\ud a_n}{\ud s}=\Gamma_ka_k(1-a_k^2\sigma^2_k),
\end{eqnarray}
which solution gives the parameters of enveloping 
gaussian anzatz for solution of FP equation, on the second stage.
From the formal point of view equations (14) are particular case of system (12).
\section{Rational Dynamics}
                                                           
After some anzatzes [15] our problems may be formulated as the systems of ordinary differential            
equations (cases 2.2 and 2.3 (system (14)) above)                                                              
\begin{eqnarray}\label{eq:pol0}                                
& & Q_i(x)\frac{\ud x_i}{\ud t}=P_i(x,t),\quad x=(x_1,..., x_n),\\
& &i=1,...,n, \quad                                                                        
 \max_i  deg \ P_i=p, \quad \max_i deg \  Q_i=q \nonumber                  
\end{eqnarray}                                                 
or a set of such systems (cases 2.1, 2.3 (full equation (13)) above) corresponding to each independent coordinate
in phase space. 
They have the fixed initial(or boundary) conditions $x_i(0)$, where $P_i, Q_i$ are not more    
than polynomial functions of dynamical variables $x_j$                                 
and  have arbitrary dependence of time. Because of time dilation                 
we can consider  only next time interval: $0\leq t\leq 1$.                      
 Let us consider a set of functions                                               
\begin{eqnarray}                                                                      
 \Phi_i(t)=x_i\frac{\ud}{\ud t}(Q_i y_i)+P_iy_i                        
\end{eqnarray}                                                  
and a set of functionals                   
\begin{eqnarray}
F_i(x)=\int_0^1\Phi_i (t)dt-Q_ix_iy_i\mid^1_0,
\end{eqnarray}
where $y_i(t) \ (y_i(0)=0)$ are dual (variational) variables.
It is obvious that the initial system  and the system
\begin{equation}\label{eq:veq}
F_i(x)=0
\end{equation}
are equivalent.
Of course, we consider such $Q_i(x)$ which do not lead to the singular
problem with $Q_i(x)$, when $t=0$ or $t=1$, i.e. $Q_i(x(0)), Q_i(x(1))\neq\infty$.

Now we consider formal expansions for $x_i, y_i$:
\begin{eqnarray}\label{eq:pol1}
x_i(t)=x_i(0)+\sum_k\lambda_i^k\varphi_k(t)\quad
y_j(t)=\sum_r \eta_j^r\varphi_r(t),
\end{eqnarray}
where $\varphi_k(t)$ are useful basis functions of  some functional
space ($L^2, L^p$, Sobolev, etc) corresponding to concrete
problem and
 because of initial conditions we need only $\varphi_k(0)=0$, $r=1,...,N, \quad i=1,...,n,$
\begin{equation}\label{eq:lambda}
\lambda=\{\lambda_i\}=\{\lambda^r_i\}=(\lambda_i^1, \lambda_i^2,...,\lambda_i^N),
\end{equation}
 where the lower index i corresponds to
expansion of dynamical variable with index i, i.e. $x_i$ and the upper index $r$
corresponds to the numbers of terms in the expansion of dynamical variables in the
formal series.
Then we put (\ref{eq:pol1}) into the functional equations (\ref{eq:veq}) and as result
we have the following reduced algebraical system
of equations on the set of unknown coefficients $\lambda_i^k$ of
expansions (\ref{eq:pol1}):
\begin{eqnarray}\label{eq:pol2}
L(Q_{ij},\lambda,\alpha_I)=M(P_{ij},\lambda,\beta_J),
\end{eqnarray}
where operators L and M are algebraization of RHS and LHS of initial problem
(\ref{eq:pol0}), where $\lambda$ (\ref{eq:lambda}) are unknowns of reduced system
of algebraical equations (RSAE)(\ref{eq:pol2}).

$Q_{ij}$ are coefficients (with possible time dependence) of LHS of initial
system of differential equations (\ref{eq:pol0}) and as consequence are coefficients
of RSAE.

 $P_{ij}$ are coefficients (with possible time dependence) of RHS
of initial system of differential equations (\ref{eq:pol0}) and as consequence
are coefficients of RSAE.
$I=(i_1,...,i_{q+2})$, $ J=(j_1,...,j_{p+1})$ are multiindexes, by which are
labelled $\alpha_I$ and $\beta_I$ are other coefficients of RSAE (\ref{eq:pol2}):
\begin{equation}\label{eq:beta}
\beta_J=\{\beta_{j_1...j_{p+1}}\}=\int\prod_{1\leq j_k\leq p+1}\varphi_{j_k},
\end{equation}
where p is the degree of polinomial operator P (\ref{eq:pol0})
\begin{equation}\label{eq:alpha}
\alpha_I=\{\alpha_{i_1}...\alpha_{i_{q+2}}\}=\sum_{i_1,...,i_{q+2}}\int
\varphi_{i_1}...\dot{\varphi_{i_s}}...\varphi_{i_{q+2}},
\end{equation}
where q is the degree of polynomial operator Q (\ref{eq:pol0}),
$i_\ell=(1,...,q+2)$, $\dot{\varphi_{i_s}}=\ud\varphi_{i_s}/\ud t$.

Now, when we solve RSAE (\ref{eq:pol2}) and determine
unknown coefficients from formal expansion (\ref{eq:pol1}) we therefore
obtain the solution of our initial problem.
It should be noted if we consider only truncated expansion (\ref{eq:pol1}) with N terms
then we have from (\ref{eq:pol2}) the system of $N\times n$ algebraical equations
with degree $\ell=max\{p,q\}$
and the degree of this algebraical system coincides
 with degree of initial differential system.
So, we have the solution of the initial nonlinear
(rational) problem  in the form
\begin{eqnarray}\label{eq:pol3}
x_i(t)=x_i(0)+\sum_{k=1}^N\lambda_i^k X_k(t),
\end{eqnarray}
where coefficients $\lambda_i^k$ are roots of the corresponding
reduced algebraical (polynomial) problem RSAE (\ref{eq:pol2}).
Consequently, we have a parametrization of solution of initial problem
by solution of reduced algebraical problem (\ref{eq:pol2}).
The first main problem is a problem of
 computations of coefficients $\alpha_I$ (\ref{eq:alpha}), $\beta_J$
(\ref{eq:beta}) of reduced algebraical
system.
These problems may be explicitly solved in wavelet approach [4]-[6].
The obtained solutions are given
in the form (\ref{eq:pol3}),
where
$X_k(t)$ are basis functions and
  $\lambda_k^i$ are roots of reduced
 system of equations.  In our case $X_k(t)$
are obtained via multiresolution expansions and represented by
 compactly supported wavelets and $\lambda_k^i$ are the roots of
corresponding general polynomial  system (\ref{eq:pol2}).
Because affine
group of translation and dilations is inside the approach, this
method resembles the action of a microscope. We have contribution to
final result from each scale of resolution from the whole
infinite scale of spaces. More exactly, the closed subspace
$V_j (j\in {\bf Z})$ corresponds to  level j of resolution, or to scale j.
We consider  a multiresolution analysis of $L^2 ({\bf R}^n)$
(of course, we may consider any different functional space)
which is a sequence of increasing closed subspaces $V_j$:
$$
...V_{-2}\subset V_{-1}\subset V_0\subset V_{1}\subset V_{2}\subset ...
$$
satisfying the following properties:
let $W_j$ be the orthonormal complement of $V_j$ with respect to $V_{j+1}: V_{j+1}=V_j\bigoplus W_j$,
then
\begin{eqnarray}
L^2({\bf R})=\overline{V_0\displaystyle\bigoplus^\infty_{j=0} W_j},
\end{eqnarray}
This functional space decomposition corresponds to exact nonlinear
eigen mode decompositions (2),(3).
It should be noted that such representations 
give the best possible localization
properties in the corresponding (phase)space/time coordinates. 
In contrast with different approaches formulae (1)-(3) do not use perturbation
technique or linearization procedures 
and represent dynamics via generalized nonlinear localized eigenmodes expansion.  
So, by using wavelet bases with their good (phase)space/time      
localization properties we can construct high-localized coherent structures in      
spa\-ti\-al\-ly\--ex\-te\-nd\-ed stochastic systems with collective behaviour.

\section{Modelling}

Resulting multiresolution/multiscale representations for solutions of equations from part 2
in the high-localized bases/eigenmodes
are demonstrated on Fig.~2--Fig.~7.
This modelling demonstrates the appearence of stable patterns formation from
high-localized coherent structures.
On Fig.~2,  Fig.~3 we present contribution to the full expansion (1)-(3) from level 1 and level 4
of decomposition (25). Figures 4, 5 show the representations for full solutions, constructed
from the first 6 eigenmodes (6 levels in formula (25)). Figures 6, 7 show stable patterns formation
based on high-localized coherent structures.

\section{Acknowledgments}

We would like to thank Professor Pisin Chen,  
Dr. Stefania Petracca and her team for
nice hospitality, help and support during Capri ICFA Workshop.

\newpage
\begin{figure}[htb]
\epsfxsize=20pc
\figurebox{25pc}{25pc}{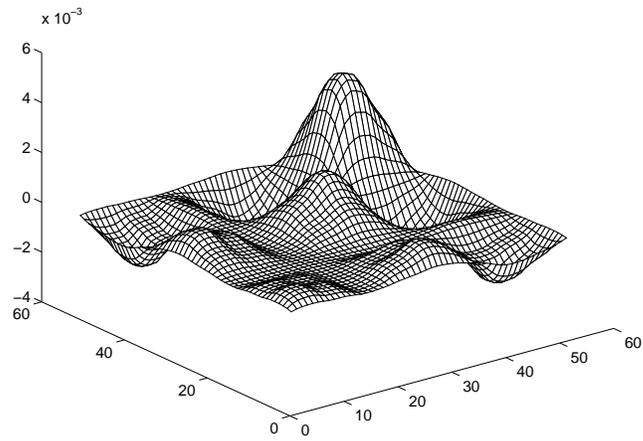}
\caption{Eigenmode of level 1}
\end{figure}

\begin{figure}[htb]
\epsfxsize=20pc
\figurebox{25pc}{25pc}{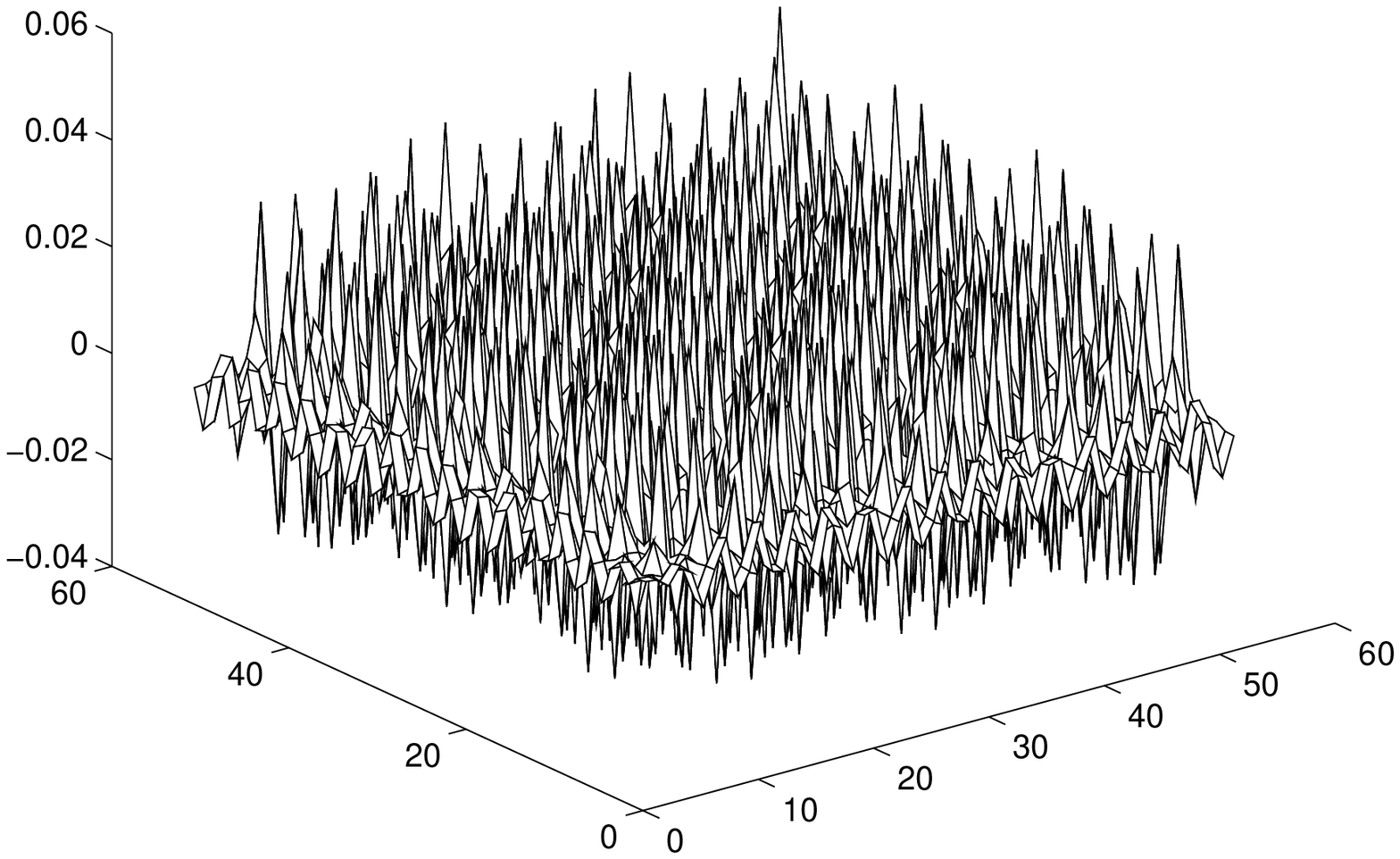}
\caption{Eigenmode of level 4}
\end{figure}

\newpage
\begin{figure}[htb]
\epsfxsize=20pc
\figurebox{25pc}{25pc}{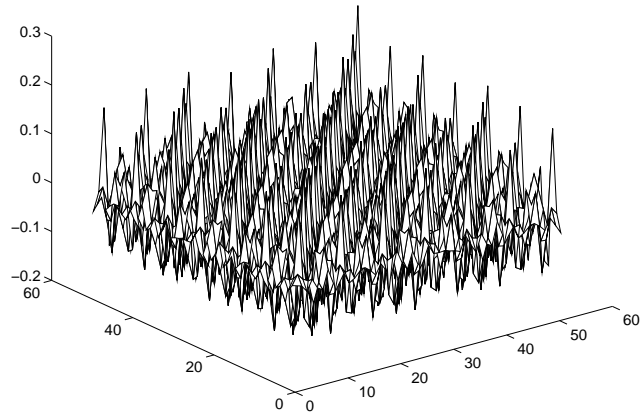}
\caption{Appearence of coherent structure}
\end{figure}

\begin{figure}[htb]
\epsfxsize=20pc
\figurebox{25pc}{25pc}{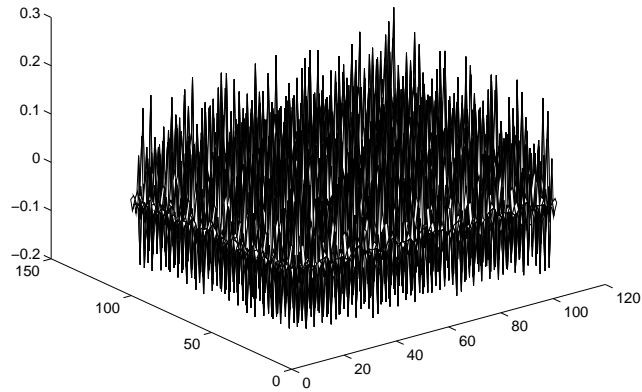}
\caption{Six-eigenmodes decomposition}
\end{figure}

\newpage
\begin{figure}[htb]
\epsfxsize=20pc
\figurebox{25pc}{25pc}{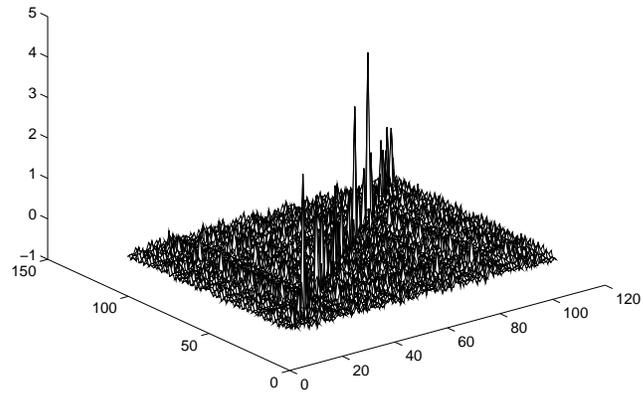}
\caption{Stable pattern 1}
\end{figure}

\begin{figure} [htb]
\epsfxsize=20pc
\figurebox{25pc}{25pc}{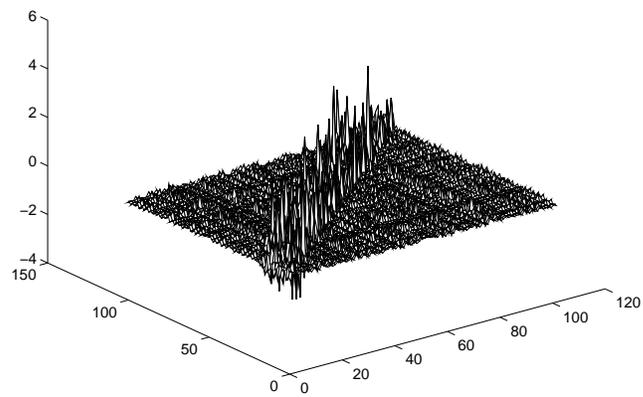}
\caption{Stable pattern 2}
\end{figure}

 \end{document}